\def\gsim{\mathrel{\rlap{\lower.6ex\hbox{$\sim$}}\raise.35ex\hbox{$>$}}}
\def\lsim{\mathrel{\rlap{\lower.6ex\hbox{$\sim$}}\raise.35ex\hbox{$<$}}}
\documentstyle[preprint,aps,eqsecnum]{revtex}
\begin{document}
\preprint{December 1993}
\title
{Singlet pairing in the double chain $t-J$ model}

\author{D.V. Khveshchenko}
\address
{Department of Physics Princeton University,\\
Princeton, NJ 08544\\
and\\
 Landau Institute for Theoretical Physics,\\
2,st.Kosygin, 117940, Moscow, Russia}

\maketitle

\begin{abstract}
\noindent
Applying the bosonization procedure to constrained
 fermions in the framework
 of the one dimensional
$t-J$ model we discuss a scenario of singlet superconductivity
in a lightly doped double chain where
all spin excitations remain gapful.
\end{abstract}
\pagebreak

Last years the problem of non-Landau Fermi liquid behavior in
 quasi-one dimensional systems again attracted a strong  interest.
This time it was stimulated by the Anderson's idea about an effectively
one
dimensional (1D) dynamics of excitations
in the normal state of layered high $T_c$ cuprates
\cite{Anderson}.

The basic problem here is a complete description
of a "dimensional crossover"
which may occur as a result of
varying coupling between  1D systems (chains)
forming a two- or three-dimensional array.
In particular, one of the main issues is whether a coherent
transport between chains establishes at arbitrary small interchain
coupling or  whether there
is a finite
 threshold resulting from the "confinement" phenomenon \cite{Anderson}.

Various weak coupling
studies of the infinite array problem don't seem to confirm
the Anderson's picture, although
one might think that the situation becomes different
at strong coupling. On the other hand systems of a finite number of chains
provide interesting examples of a peculiar
behavior in an "intermediate dimenion".
In analogy with a purely 1D  case one might
expect that these also allow a consistent strong coupling treatment.

Moreover these models can also describe properties
of such real materials as $(VO)_{2}P_{2}O_{7}$ or
 $Sr_2Cu_4O_6$ which contain weakly coupled
metal-oxide-metal double chain ladders.
It was also pointed out in \cite{RGS} that  higher stoichiometric compounds
$Sr_{n-1}Cu_{n+1}O_{2n}$ provide a physical realization
of weakly coupled $N$-chain ladders.

A proper Hamiltonian of a strongly correlated double chain
is that of the $t-J$ model with antiferromanetic spin exchanges \cite{RGS}:
\begin{eqnarray}
H=-\sum_{i,\sigma}
(t\sum_{f}c^{\dagger}_{i,\sigma,f}c_{i+1,\sigma,f}+
t_{\perp}c^{\dagger}_{i,\sigma,u}c_{i,\sigma,d}+h.c.)+\nonumber\\
+\sum_{i}(J{\vec S}_{i,f}{\vec S}_{i+1,f}+J_{\perp}
{\vec S}_{i,u}{\vec S}_{i,d}),
\end{eqnarray}
where  $f=u,d$ is the chain index.
The Hamiltonian (1) has to be complemented by the no double occupancy
constraint
\begin{equation}
\sum_{\sigma}c^{\dagger}_{i,\sigma,f}c_{i,\sigma,f}\leq 1
\end{equation}
Accordingly, a  weak coupling regime can be studied in the framework
of the Hubbard model.

At half filling the Hamiltonian (1) describes $S=1/2$ double chain
Heisenberg model. Available numerical results \cite{DRS},
\cite{NWS}
 as well as a mean field
analysis using the Gutzwiller projection
\cite{SRZ} indicate that in contrast to the case of a single chain
there are no gapless spin excitations in a double chain.
Moreover the spin gap appears to be robust against doping and survives in some
range around  half filling.
These observations are in agreement with the conjecture \cite{RGS}
that a lightly doped
double chain system becomes a singlet superconductor of a modified
d-wave type.

Recent weak coupling renormalization group (RG)
studies of the double-chain Hubbard model did
reveal some spin gapful fixed points characterized  by an inhanced
singlet pairing in both cases $U<t\sim t_\perp$ \cite{FPT}
and $t_\perp <U<t$
\cite{KR}.
An inhancement and a power-law decay of pairing correlations were also
convincingly shown numerically \cite{NWS}.

To clarify the essence of the double chain physics
 it is worthwhile to review  properties of the
single chain $t-J$ model.

In the region of ratios $J/t\lsim 1$ the
model can be only found in the so-called Tomonaga- Luttinger (TL)
regime which corresponds
to both gapless spin and charge excitations \cite{OLSA},\cite{HM}.
It is customary to describe the TL  behavior in terms of spin and charge
correlation exponents $K_{s}$ and
$K_{c}$.

The spin exponent $K_{s}$ equals to unity
everywhere in the TL regime while $K_{c}$
gradually increases from the value
1/2 which it reaches at $J=0$ and any density $\rho$
as well as at $\rho\rightarrow 1$
and arbirary $J/t$ as $J/t$ increases or $\rho$ gets smaller.
The TL regime persists up to  $J/t\approx2.5$ where a
spin gap with strong pairing correlations occurs at
 small enough fermion density ($\rho\lsim 1/3$).
In fact, one can understand the
occurence of the region of attraction at small $\rho$ as resulting from
the  existence of a two-particle bound state at zero density.
A finite threshold in the attraction
strength follows from  vanishing of the bound state wave function
at zero separation due to the no-double occupancy
constraint.

   On the other hand in the
regime of strong correlation at $\rho$ close to unity
one can argue that gapless spin
fluctuations drive  couplings of the charge sector
to the repulsive region and to get
an effective attraction $(K_{c}>1)$
one has to exceed some threshold value of $J/t$.

However if both $\rho$ and  the critical value of $J/t$
are large then the attraction of charges
actually leads to the phase separation rather than to the real
superconducting pairing.
Even the inclusion of the short-range repulsion which is supposed
 to postpone
the onset of phase separation to higher $J/t$ doesn't  extend the
region of singlet pairing \cite{TTRRD}.

If, on the contrary, spin fluctuations are gapful
then these may not renormalize charge couplings
significantly.
Therefore the  charge correlation exponent
may not receive its basic contribution leading to $K_{c}<1$ at
small $J/t$ and the mechanism
of attraction may work without any threshold in J/t. Various
 possibilities to get
such a behavior by means of the frustrating spin exchange interactions
in a single chain were
 considered in \cite{I},
\cite{OLR}.

Coupling between Luttinger
chains provides an alternative way to produce a spin gap
 favoring attraction between charges.

Preceding analytical studies of the problem in the framework of the bosonized
"g-ology" \cite{FPT},\cite{FL},\cite{KR} were restricted on the case
of weak coupling.

 In the present
paper we find additional arguments in favor of the above scenario
 by using
a bosonic representation of the $t-J$ model which is an adequate tool
to study a strong coupling behavior.

Although the method of
bosonization is  conventionally applied to 1D  weakly
interacting fermions
it might be possible to formulate a consistent procedure
for the opposite limit when the interaction is extremely strong, that is
for the case of constrained fermions.
Various versions of the bosonization
procedure in the  framework of the $t-J$ model were
 discussed in the literature \cite{SB},\cite{WSTS}.
Recently a modification of the approach proposed in \cite{WSTS}
was shown to give correct exponents for the one- and two-particle
correlation functions as well a good approximation
for the  energy spectrum of the single
chain $t-J$ model \cite{FSY}.

According to the method of \cite{FSY} the constrained fermion operator
$c_{i,\sigma}$ can be
represented as a product of a spinless fermion $\Psi_{i}$
and a spin one-half operator (hard-core boson) $S_{i}^{\pm}$:
\begin{equation}
c_{i\uparrow}=P_{i}\Psi_{i}S^{-}_{i}P_{i}^{\dagger},
{}~~~~~~~c_{i\downarrow}=P_{i}\Psi_{i}S^{+}_{i}P_{i}^{\dagger}
\end{equation}
where the projection operator $P_{i}$
reduces the space of four on-site states ($|hole>\otimes |spin>=
|1,\uparrow>,|1,\downarrow>,
|0,\uparrow>,|0,\downarrow>$) to the physical Hilbert space formed by
the set $|0>, |\uparrow>,|\downarrow>$.

In turn, the spin one-half operator $S_{i}^{\pm}$ can be expressed in terms
of a spinless Jordan-Wigner fermion
\begin{equation}
S^{+}_{i}=\chi_{i}^{\dagger}
 e^{i\pi\sum_{j<i}\chi^{\dagger}_{j}\chi_{j}}~~~~~~~
S^{-}_{i}=\chi_{i} e^{-i\pi\sum_{j<i}\chi^{\dagger}_{j}\chi_{j}}~~~~~~~
S^{z}_{i}=\chi^{\dagger}_{i}\chi_{i}-{1\over 2}
\end{equation}

The authors of \cite{FSY} also argued that  the
local constraint (2) and the sum rule for the constrained  fermions
$\sum_{\sigma}\int_{0}^{\infty}{d\omega\over {2\pi}}Im<\{c_{i,\sigma}^{\dagger}
(\omega),c_{i,\sigma}(-\omega)\}>=1+\delta$
(where $\delta$ is doping) are
obeyed even in the approximation which discards
the projector $P_{i}$. Therefore  this approximation which reportedly yields
correct exponents of correlation functions and momentum distribution
was supposed to provide
a  better account of fermion correlations than other approaches
which don't treat hard-core
condition properly.

With the neglect of projectors $P_i$
the representation (3) allows one to rewrite (1) in the form
\begin{eqnarray}
H=-\sum_{i,\sigma}
(t\sum_{f}\Psi^{\dagger}_{i,f}\Psi_{i+1,f}
(S^{+}_{i,f}S^{-}_{i+1,f}+S^{-}_{i,f}S^{+}_{i+1,f})
+t_{\perp}\Psi^{\dagger}_{i,u}\Psi_{i,d}
(S^{+}_{i,u}S^{-}_{i,d}+S^{-}_{i,u}S^{+}_{i,d})+h.c.)+\nonumber\\
+\sum_{i}(Jn_{i,f}{\vec S}_{i,f}{\vec S}_{i+1,f}n_{i+1,f}+J_{\perp}
n_{i,u}{\vec S}_{i,u}{\vec S}_{i,d}n_{i,d})-\mu\sum_{i,f}
\Psi^{\dagger}_{i,f}\Psi_{i,f}
\end{eqnarray}
with a local charge density defined as
$n_{i,f}=\Psi^{\dagger}_{i,f}\Psi_{i,f}$ and a local spin
${\vec S}_{i,f}$ given in terms of $\chi_{i,f}$ according to (4).

To get a bosonic form of the Hamiltonian (5) one can use the representation
\begin{equation}
\Psi_{i,f}\sim \sum_{\mu=R,L}e^{i\mu\pi(1-\delta)x+i\mu \phi_{c}^{f}(x)+i
\theta_{c}^{f}(x)}
{}~~~~~~~~~\chi_{i,f}\sim\sum_{\mu=R,L}e^{
i\mu\pi x+i\mu \phi_{s}^{f}(x)+\theta_{s}^{f}(x)}
\end{equation}
Keeping the most relevant operators
in the continuous limit we obtain the bosonized Hamiltonian density
\begin{eqnarray}
H_B={1\over 2}\sum_{\pm}\{v^{\pm}_{c}((\partial\theta_{c}^{\pm})^2+
(\partial\phi_{c}^{\pm})^2
+\partial\phi_{c}^{\pm}\partial\phi_{s}^{\pm}
-\partial\theta_{c}^{\pm}\partial\theta_{s}^{\pm})+\nonumber\\
+v_{s}((\partial\theta_{s}^{\pm})^2+
(\partial\phi_{s}^{\pm})^2)
+J(1-\delta+{1\over {\sqrt{2}\pi}}(\partial\phi_{c}^{+}\pm
\partial\phi_{c}^{-}))^{2}
\cos 2(\phi^{-}_{s}\pm\phi^{+}_{s})\}+\nonumber\\
+t_{\perp} (\cos\phi^{-}_{c}+\cos(\phi^{+}_{c}+2\delta x)
\cos\phi^{+}_{s})
\cos\theta^{-}_{c}\cos\theta^{-}_{s}+\nonumber\\
+
J_{\perp}
(1-\delta-{1\over {\sqrt{2}\pi}}(\partial\phi_{c}^{+}+\partial\phi_{c}^{-}))
(1-\delta-{1\over {\sqrt{2}\pi}}(\partial\phi_{c}^{+}-\partial\phi_{c}^{-}))
(\cos 2\phi^{+}_{s}+\cos 2\phi^{-}_{s}+\cos \theta^{-}_{s})
\end{eqnarray}
where $\phi_{c,s}^{\pm}={1\over {\sqrt 2}}(\phi_{c,s}^{u}\pm
\phi_{c,s}^{d})$ and
$\theta_{c,s}^{\pm}={1\over {\sqrt 2}}(\theta_{c,s}^{u}\pm
\theta_{c,s}^{d})$. As usual, the values of  correlation exponents
$K_{c,s}^{\pm}$ can be  affected by short-wavelength
renormalizations. The bare
 velocities of charge and
spin excitations are given by the formulae:
$v_c =2t\sin\pi\delta$ and
$v_s =2J((1-\delta)^2-({\sin\pi\delta\over \pi})^2)+4\pi t\sin\pi\delta $
although these can be altered too.

We note that the expression (7) can be also obtained by using the $CP^1$
coherent state representation first introduced in \cite{W} and applying
a somewhat different  bosonization scheme discussed in \cite{SB}.
We believe that it verifies a neglect of the projecting operators when
deriving a relevant part of the continuous bosonic Hamiltonian.

 The first three terms in (7) can be recognised as the Hamiltonian
of the charge carrying spinless fermion coupled to the Abelian gauge field
$A_{\mu}=\epsilon_{\mu\nu}\partial_{\nu} \phi_{s}
=(\partial\phi_{s},\partial\theta_{s})$
which describes a surrounding spin
background.
Notice that once the constraint was explicitly resolved one obtains only
two independent fields instead of three (spinon, holon and a gauge
field) appearing in mean field studies of the $t-J$ model.

In the case of a single chain the basic TL phase
 of the $t-J$ model corresponds to
 both  $\phi_{c}$ and $\phi_{s}$ being gapless.
Scaling dimensions of operators of the form
$\cos\beta\phi\cos\beta^{\prime}\theta$ can be estimated by means of
 the formula
\begin{equation}
\Delta={1\over 4}(K\beta^2 +{\beta^{\prime 2}\over K})
\end{equation}

At $\delta =0$ (7) becomes equivalent to the bosonized
Heisenberg double chain due to
 the effective freezing of the charge degrees of freedom.
 The resulting expression essentially  coincides
with the relevant part of the one obtained
in \cite{SM} where a more general $XXZ$ symmetrical case was considered.

It was arued in \cite{SM}  that at least one of the operators
$\cos 2\phi_{s}^{-}$ and  $\cos \theta_{s}^{-}$
 appears to be relevant and drives the "-" spin sector toward a strong
coupling regime where either $\phi_{s}^{-}$ or $\theta_{s}^{-}$ gets locked
and  a corresponding cosine acquires a nonzero expectation value.

Additionally, at
$K_{s}^{+}<1$ the "+" sector gets to a strong coupling regime
 where $\phi_{s}^{+}$ is locked.
Therefore in general there are four possible phases
with  finite  $v_{s}$ which
can be identified as follows $(\eta=2/K_{s}^{+})$:\\
-Both $\phi_{s}^{-}, \phi_{s}^{+}$ are locked-
 the antiferromagnetically  ordered state
;\\
- Both $\theta_{s}^{-}$ and $ \phi_{s}^{+}$ are locked - singlet state
(all spin excitations are gapful);\\
-$\theta_{s}^{-}$ is locked -  the XY-type phase
characterised by correlations:
\begin{eqnarray}
<S^{z}_{f}(x)S^{z}_{f}(0)>\sim {\frac{1}{x^2}}+(-1)^x e^{-x}\nonumber\\
<S^{+}_{f}(x)S^{-}_{f}(0)>\sim (-1)^x {\frac{1}{x^{\eta}}}+e^{-x}
\end{eqnarray}
- $\phi_{s}^{-}$ is locked - another gapless phase
having  different spin correlations:
\begin{eqnarray}
<S^{z}_{f}(x)S^{z}_{f}(0)>
\sim {\frac{1}{x^2}}+(-1)^x  {\frac{1}{x^{1/\eta}}}\nonumber\\
<S^{+}_{f}(x)S^{-}_{f}(0)> \sim e^{-x}
\end{eqnarray}
All these states besides the last one
were argued in \cite{SM} to appear on an extended phase diagram
of the double chain  $XXZ$
symmetrical (Heisenberg-Ising) spin model.
The analysis carried out in \cite{SM} leads to the conclusion
 that at $J_{\perp}>-{1\over 4}J^{z}$
the gapless line $J_\perp =0$
becomes unstable against arbitrary small $J_\perp$ of any sign.

It has to be noticed that at $J_{\perp}<0$ and
$J_{\perp}>0$ the nature of the singlet
ground  state
is quite different. In the former case every pair of spins on one rung of
the ladder tends to form a $S=1$ state and the
system effectively behaves similar to $S=1$
Heisenberg chain while in the latter case spin pairs couple preferably
into singlets  which then form a "dimer liquid".

According to the phase diagram
proposed in \cite{SM}
 the spin isotropic poin at $J_\perp >0$ is located deeply
 inside the gapful "dimer liquid" phase
with both $\theta_{s}^{-}, \phi_{s}^{+}$ being locked.
It agrees with numerical \cite{DRS},\cite{NWS} and mean field \cite{SRZ}
results.

We also note that in a general case of $N$-chain ladders
one may expect that a spin gap is present at even $N$ only.
To see that one can apply arguments due to Haldane
\cite{H}. In \cite{H} a topological term governing a longwavelength dynamics
of the 2D lattice Heisenberg
model was found in the form $\sum_{y}(-1)^y Q_{y}(x,t)$,
where $Q_{y}(x,t)$ is a topological
$\theta$-term appearing in a  purely 1D
 case and distinguishing between
integer and half-odd integer spins \cite{A}.
In the 2D case with periodic boundary conditions this sum is equal to zero
which means the absence of a 2D counterpart of the 1D $\theta$-term.
However applying this formula to the finite width strip
one can  see that for odd $N$  the above sum doesn't vanish
and therefore an effectively 1D longwavelength
dynamics remains gapless. In contrast, for even $N$ the gap
survives and then scales as $\Delta(N)\sim \exp(-N)$.

On the basis of the results obtained in
\cite{DRS}, \cite{NWS}, \cite{SRZ}
we assume that at small doping
the  hopping terms in (7)  can be treated as perturbations
which do not destroy the gapful spin state.
Then we get the effective Hamiltonian describing a charge
dynamics in the spin gap
state
\begin{equation}
H_{c}={1\over 2}\sum_{\pm}v^{\pm}_{c}(K_{c}^{\pm}
(\partial\theta_{c}^{\pm})^2+{1\over K_{c}^{\pm}}
(\partial\phi_{c}^{\pm})^2)+{\bar t}_{\perp}\cos\phi_{c}^{-}
\cos\theta_{c}^{-}
\end{equation}
where ${\bar t}_{\perp}=t_{\perp}<\cos\theta_{s}^{-}>$.
Charge correlation exponents which can be easily
read off from (5)
are given by the formulae
\begin{equation}
K_{c}^{\pm}=(1+{J\over {\pi t}}
<{\vec S}_{f}{\vec S}_{f}>\pm {J_{\perp}\over {\pi t}}
<{\vec S}_{u}{\vec S}_{d}>)^{-1/2}
\end{equation}
Due to the short-range antiferromagnetic order we encounter the case
of $K_{c}^{+}>1$  while $K_{c}^{-}-1$ can be, in principal,  of both signs.

 The physical origin of attraction between charges in the paramagnetic
spin gap state with a short range antiferromagnetic order
 can be understood on very general
grounds
\cite{FS}. A straightforward  manifestation of this
phenomenon  in the framework of the $t-J$ model
is a negative sign in front of the product of charge densities
$n_{f}
n_{f^{\prime}}$ staying in (5) if $<{\vec S}_{f}{\vec S}_{f^{\prime}}><0$.

A progressive understanding of the spinless double chain problem
 \cite{KLN},\cite{Y},\cite{NLK}
shows
 that despite of the possibly vanishing  of the single particle
hopping the "-" charge sector of the system always evolves to
the strong coupling regime due to the development of either coherent
particle-particle or particle-hole pair hopping.
This phenomenon  was previously discussed  by many authors
in the context of quasi-one dimensional conductors \cite{S}.

For the account  of these processes triggered by the single
particle hopping
the charge Hamiltonian (11) has to be supplemented by the extra terms
\begin{equation}
\delta H_{c}=
g_{ph}\cos 2\phi_{c}^{-}+g_{pp}\cos 2\theta_{c}^{-}
\end{equation}
  generated in the
course of renormalization. At small ${\bar t}_{\perp}/t$
a conventional RG procedure applied to
 the extended
Hamiltonian (11,13) leads to the system of equations describing
a renormalization flow in the "-" charge sector \cite{NLK},\cite{KR}
($\xi$ is a scaling variable):
\begin{eqnarray}
\frac{d g_{ph}}{d\xi}
=2(1-K_{c}^{-})g_{ph} +{{\bar t}_{\perp}^2}
(K_{c}^{-}-{1\over K_{c}^{-}})\nonumber\\
{{d g_{pp}}\over {d\xi}}
=2(1-{1\over K_{c}^{-}})g_{pp} +{{\bar t}_{\perp}^2}
({1\over K_{c}^{-}}-K_{c}^{-})
\nonumber\\
{{d\log K_{c}^{-}}\over {d\xi}}={1\over 2}(-K_{c}^{-}g_{ph}^2+
{1\over K_{c}^{-}}g_{pp}^2)
\end{eqnarray}
The analysis of the solutions of (14) first performed in \cite{NLK}
shows that depending on the sign of
$g_{-}=g_{ph}-g_{pp}$ either $\cos 2\phi_{c}^{-}$ (at $g_{-}>0$) or
$\cos 2\theta_{c}^{-}$ (at $g_{-}<0$) acquires a nonzero
expectation value. The asymptotic behaviors of the correlation exponent
in the two cases
are
$K_{c}^{-}(\xi)\rightarrow 0$ and
$K_{c}^{-}(\xi)\rightarrow \infty$
respectively. By considering four possible
 order parameters for the spinless case
\begin{eqnarray}
{ {CDW}}_{+}
\sim \cos(\phi^{+}_{c}+\phi^{-}_{c})~~~~~~~~
{ {CDW}}_{-}
\sim \cos(\phi^{+}_{c}+\theta^{-}_{c})\nonumber\\
{{SS}}_{+}
\sim \cos(\theta^{+}_{c}+\theta^{-}_{c})~~~~~~~~
{ {SS}}_{-}
\sim \cos(\theta^{+}_{c}+\phi^{-}_{c})
\end{eqnarray}
 one concludes that at $g_{-}>0
$ the competing  types of ordering are (intrachain)
$CDW_{+}$ and (interchain) $SS_{-}$ while at
$g_{-}<0$ the relevant orderings are $CDW_{-}$ and $SS_{+}$.
In turn, the result of the
competition between them depends on the sign of $g_{+}=g_{ph}+g_{pp}$.

By mapping the Hamiltonian (11,13)
onto the spin $S=1/2$ chain in an external magnetic field the authors
of \cite{NLK} also argued
that the above statements hold
at strong coupling too.

Considering different order parameters relevant for
fermions with spin
\begin{eqnarray}
CDW_{+}=\sum c^{f\dagger}_{\mu\sigma}c^{f}_{-\mu,\sigma}\sim
\cos(\phi^{+}_{c}+\phi^{-}_{c})\cos(\phi^{+}_{s}+\phi^{-}_{s})\nonumber\\
CDW_{-}=\sum c^{f\dagger}_{\mu\sigma}c^{-f}_{-\mu,\sigma}\sim
\cos(\phi^{+}_{c}+\theta^{-}_{c})\cos(\phi^{+}_{s}+\theta^{-}_{s})\nonumber\\
SDW_{+}=\sum c^{f\dagger}_{\mu\sigma}c^{f}_{-\mu,-\sigma}\sim
\cos(\phi^{+}_{c}+\phi^{-}_{c})\cos(\theta^{+}_{s}+\theta^{-}_{s})\nonumber\\
SDW_{-}=\sum c^{f\dagger}_{\mu\sigma}c^{-f}_{-\mu,-\sigma}\sim
\cos(\phi^{+}_{c}+\theta^{-}_{c})\cos(\theta^{+}_{s}+\phi^{-}_{s})\nonumber\\
SS_{+}=\sum\sigma c^{f}_{\mu\sigma}c^{f}_{-\mu,-\sigma}\sim
\cos(\theta^{+}_{c}+\theta^{-}_{c})\sin(\phi^{+}_{s}+\phi^{-}_{s})\nonumber\\
SS_{-}=\sum\sigma c^{f}_{\mu\sigma}c^{-f}_{-\mu,-\sigma}\sim
\cos(\theta^{+}_{c}+\phi^{-}_{c})\sin(\phi^{+}_{s}+\theta^{-}_{s})\nonumber\\
TS_{+}=\sum\sigma c^{f}_{\mu\sigma}c^{f}_{-\mu,\sigma}\sim
\cos(\theta^{+}_{c}+
\theta^{-}_{c})\sin(\theta^{+}_{s}+\theta^{-}_{s})\nonumber\\
TS_{-}=\sum\sigma c^{f}_{\mu\sigma}c^{f}_{-\mu,\sigma}\sim
\cos(\theta^{+}_{c}+\phi^{-}_{c})\sin(\theta^{+}_{s}+\phi^{-}_{s}),
\end{eqnarray}
we observe that in contrast to the spinless case,
any paramagnetic spin gapful
 state has  only two relevant types of orderings
which are $SS_{-}$ and $CDW_{-}$ containing fields $\theta^{-}_{s}$
and $\phi^{+}_{s}$ \cite{KR}.

Examining the divergencies of
corresponding response functions with the use of (8) and (12)
we obtain that at $J_\perp <{\vec S}_{u}{\vec S}_{d}> <0$
the interchain singlet pairing $SS_{-}$ appears to be
the leading instability  while in the opposite case
 $J_\perp <{\vec S}_{u}{\vec S}_{d}> >0$ the ground state is the
"flux phase" $CDW_{-}$.

 The latter state is characterised
by the  commensurate with density "flux" $\Phi=2k_{F}$
defined as a circulation
of a phase of the on-rung order parameter
 $<u^{\dagger}_{i}d_{i}+d^{\dagger}_{i}u_{i}>$  through a plaquette
formed by two adjacent rungs of the ladder.
In the case of spinless fermions this type of ordering
called the "Orbital Antiferromagnet" was first discovered in \cite{N} as a
counterpart of  2D flux states.

It is also instructive to express the above order parameters in terms
of the hybridized
states corresponding to the  mean field "bonding" and "antibonding"  bands
$B,A={1\over {\sqrt 2}}(u\pm d)$:
\begin{eqnarray}
CDW_{-}=\sum_{\sigma}
A^{\dagger}_{R\sigma}A_{L\sigma}-B^{\dagger}_{R\sigma}B_{L\sigma}\nonumber\\
SS_{-}=\sum_{\sigma}
A^{\dagger}_{R\sigma}A^{\dagger}_{L,-\sigma}
-B^{\dagger}_{R\sigma}B_{L,-\sigma}^{\dagger}
\end{eqnarray}
Considering the distribution of signs of the order parameter
 $SS_{-}$ on the "four-point Fermi surface"
$({\vec k}=(k_F ,0),(-k_F ,0),(k_F ,\pi),(-k_F ,\pi))$
we observe that it corresponds to the "d-wave" type pairing.
One might expect
that in a
two-dimensional array of weakly coupled double chains with a continuous
 Fermi surface this
type of ordering does transform into an ordinary d-wave pairing.

It follows from the preceding discussion that
both instabilities develop without a threshold in $J/t$ or $J_{\perp}/t$.
Note that this statement is in agreement
with the results of the weak coupling analysis
of the double chain Hubbard model \cite{KR}.

We also add that in the antiferromagnetically  ordered state
where all spin excitations acquire Ising gaps, the only relevant order
parameters could
be $SS_{+}$ and $CDW_{+}$ containing  fields
$\phi_{s}^{-}$ and  $\phi_{s}^{+}$.
However at all $J<{\vec S}_{f}{\vec S}_{f}>~<J_{\perp}^2/t$
the intrachain pairing $SS_{+}$ is always favored.

In summary, in the present paper we applied the method of
bosonization to
constrained
fermions in the context of the double chain $t-J$ model.
As a result, we found  further arguments supporting
the recently proposed scenario of singlet superconductivity in the spin gap
state of the double chain problem \cite{RGS}.

Our analysis was based on the assumption that a small doping doesn't
destroy the spin gap and a spin dynamics remains essentially the same as
in the insulating case. This conjecture is strongly supported
by the results of numerical studies
\cite{DRS},\cite{NWS}, the Gutzwiller projected  mean field  \cite{SRZ}
and weak coupling "g-ology" \cite{FPT},\cite{KR}.

We also want to stress that  our conclusions contradict
with a recent claim \cite{ZFL}  about
an existence of the strong
coupling fixed point where some spin excitations remain
gapless. The authors of \cite{ZFL} considered the double chain $t-J$ model
without an interchain spin exchange ($J_\perp =0$).
Then on the bare level their Hamiltonian can be assigned to the universality
class of the purely
forward scattering  model considered in \cite{FL}.
Indeed,
in this special case
the only field
becoming massive
is $\theta^{-}_{s}$. In principal, it can't be ruled out that for some specific
double chain models only one of two spin fields becomes gapful.
However an investigation of the double chain
Hubbard model \cite{FPT},\cite{KR} demonstrates that
the presence of the interchain
one particle hopping is already sufficient to generate
the
antiferromagnetic spin exchange term
with $J_\perp \sim \frac{t^{2}_{\perp}}{max(t,U)}$
which will eventually make all spin modes gapful. We believe that it
is a general feature
of  spin isotropic
 models of strongly correlated fermions on double chains.

The author is indebted to Profs. T.M.Rice and F.D.M.Haldane for
valuable discussions of these and related issues.
This work was  supported by the NSF Grant.

\pagebreak

\end{document}